\documentclass[pra,twocolumn,showpacs,twoside,floatfix]{revtex4}
\usepackage{graphicx}
\usepackage{amsmath}
\usepackage{placeins}
\usepackage{hyperref}

\newcommand{\uo}{\mu_o}
\newcommand{\TU}{T_U}
\newcommand{\muU}{\mu_U}

\newcommand{\uloc}{\mu_\mathrm{loc}}
\newcommand{\Ui}{U_\mathrm{int}}
\newcommand{\otrap}{\omega_\mathrm{trap}}

\newcommand{\Vh}{V_\mathrm{harm}}

\newcommand{\eq}{\begin{eqnarray}}
\newcommand{\en}{\end{eqnarray}}

\newcommand{\rcut}{r_\mathrm{cut}}

\newcommand{\state}[2]{$\left|#1\right>_{#2}$}

\def\ket#1{\mathinner{|{#1}\rangle}}

\begin{document}

\title{Shaking the entropy out of a lattice: atomic filtering by vibrational excitations}

\author{Malte C. Tichy}
\author{Klaus M{\o}lmer}
\author{Jacob F. Sherson}
\affiliation{
Lundbeck Foundation Theoretical Center for Quantum System Research, \\
AU Ideas Center for Community Driven Research, \\
Department of Physics and Astronomy, University of Aarhus, DK-8000 Aarhus C, Denmark}

\date{\today}


\begin{abstract}
We present a simple and efficient scheme to reduce  atom-number fluctuations in optical lattices.  The interaction-energy difference for atoms in different vibrational states is used to remove excess atomic occupation. The remaining vacant sites are then filled with atoms by merging adjacent wells, for which we implement a protocol that circumvents the constraints of unitarity. The preparation of large regions with precisely one atom per lattice site is discussed for both bosons and fermions. The resulting low-entropy Mott-insulating states may serve as high-fidelity register states for quantum computing and as a starting point for investigations of many-body physics.
\end{abstract}

\maketitle

\section{Introduction}
An optical lattice with single-atom occupancy of each site is the ideal low-entropy starting point for quantum computation ~\cite{Deutsch99:qGatesOptLatt,Calarco00:Qgates,meschedePRL04:Qregister,rolstonPRA06:singleSite,Negretti:2011zr} and for the simulation of condensed matter systems \cite{Lewenstein07:simulationReview,Bloch:2012ys}. Even at the lowest temperatures achieved to date, however, defects, {\it i.e.}~vacant or multiply occupied sites, are still common~\cite{Greiner10:science,Sherson10:MInature,Bakr:2011fk}  and jeopardize the aforementioned applications: For these, exceptionally low temperatures are required, and neither vacancies nor multiple occupation of lattice sites can be tolerated. Cooling and thermometry schemes beyond the nanokelvin range are therefore being developed \cite{0034-4885-74-5-054401}. The purification of the state is  a challenge: For an ideal low-entropy Mott insulator with single-atom occupancy of lattice sites, one needs to reliably eliminate multiple occupations using number filtering, and, subsequently, fill vacant lattice sites with atoms from a reservoir.

Multiple occupations -- or, for fermions, occupation of higher excited states \cite{PhysRevA.82.011610} -- can be dealt with by global manipulation~\cite{0953-4075-43-13-131001,Rabl03,Popp06:gndStateCooling}, using the difference in the interaction energy between internal states to make  transition frequencies density-dependent. In this way higher single well occupation numbers can be filtered to unity.

To avoid vacant sites, one can use single-site resolution to first measure the occupation of the lattice in a site-resolved manner and to subsequently rearrange atoms using spin-dependent lattices, site-selective spin-flips and spin-selective shifts \cite{Weiss04:zeroEntropy}. A crucial requirement of this proposal is the ability to recool atoms to their vibrational ground state, which despite recent advances~\cite{Li2012} still remains an unsolved challenge. Alternatively, one can utilize global many-body dynamics~\cite{Popp06:gndStateCooling,Ho09:cooling,Kollath09:cooling} where entropy is removed followed by one or more periods of  global equilibration.

While the combination of filtering and tunneling operations ideally prepares a lattice with unit occupancy of each well, the difference in interaction energy between different hyperfine states may be too small~\cite{Ketterle06:MIclockShift} and the thermalization may be too slow for the procedure to compete with the heating under realistic experimental conditions. On the other hand, the dependence of the interaction energy between atoms on their spatial overlap implies a considerable difference in the interaction energy for different vibrational states. Experimentally, the removal of excess atoms was recently performed using parametric lattice modulation by exploiting different vibrational states \cite{Bakr:2011fk}. The problem of vacancies has not been tackled experimentally so far.

In this Article we introduce a combined purification protocol for number-filtering and filling of vacant sites, for a thermal Mott insulator. We achieve occupation-number-dependent transfer that does not rely on the usually small natural difference in interaction strength for different hyperfine states, but instead we use the interaction shifts of different vibrational states~\cite{Bakr:2011fk}.

We will use these interaction shifts for \emph{(i)} number filtering, where we remove excess atoms from the wells (Section \ref{numberfiltering}) and for \emph{(ii)} filling of vacancies by merging, where we merge each lattice well with two auxiliary traps in a unitary way to establish unit filling of each well (Section \ref{sec:merging}).  We then apply our results to the experimentally relevant case of a shallow external confinement with a non-uniform thermal distribution of atoms and vacancies, in Section \ref{sec:applass}. We analyze how the central domain approaches the perfect state, and how the size of this perfect domain grows with successive merging steps. Section \ref{sec:conclusions} summarizes our results.

\section{Removing excess atoms: Number filtering} \label{numberfiltering}
As starting point we consider a system where only a particular hyperfine state ($\alpha$) and the vibrational ground state are populated, but with fluctuating occupation from site to site. In our numerical examples below, the starting point will be low temperature approximations to the Mott-Insulating (MI) state where the vacancy probability is a function of the interaction strength, the temperature and the chemical potential (see Eq.~(\ref{eq:Pn})).

\begin{figure}[h]
\includegraphics[width=\linewidth]{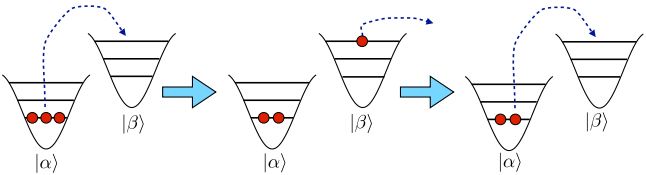}
\caption{(color online) Filtering step, where the quantum state with $n=3$ vibrational ground state atoms in ($\alpha$) is resonantly coupled by Raman laser beams to the state with one vibrationally excited atom in ($\beta$). This atom is subsequently removed. The sequence is then repeated for $n=2$.}
\label{fig:Scheme}
\end{figure}

To filter the population on all sites so that any higher occupancy is reduced to unity, we propose to drive a Raman transition from the vibrational ground state of one hyperfine state ($\alpha$) to the second vibrationally excited level of another hyperfine state ($\beta$). Since the atomic population is transferred between states with different spatial distributions~(see Fig.~\ref{fig:Scheme}), the  interaction energy in the initial and final state depends significantly on the occupation of the states. A pair of atoms residing in vibrational states $|\nu \rangle$ and $|\mu \rangle$ have an interaction energy
\eq U_{\nu,\mu}  = (2-\delta_{\nu,\mu}) \frac{4 \pi \hbar^2 a_{\text{sc}}}{m} \int |\psi_\nu(\vec{r})|^2  |\psi_\mu(\vec{r})|^2 \text{d}^3\vec{r}, \en
where the constant of proportionality is related to the s-wave scattering length $a_{\text{sc}}$ and includes a factor accounting for exchange \cite{Schneider:2009fk}. For simplicity, we ignore in the following the difference between the internal state interaction strengths (for $^{87}$Rb $U_{\alpha\beta}\sim0.98 U_{\alpha\alpha}$~\cite{Ketterle06:MIclockShift}), and we assume anisotropic harmonic oscillator wave functions for the lowest vibrational states. In the harmonic approximation we find the following relative coupling strengths for the lowest levels (states are labeled  by the 1D vibrational quantum number in the direction of excitation),
\eq U_{\nu,\mu}=U_{00} \left(\begin{tabular}{cccc}
1 & 1 & 0.75 & \dots \\
1 & 0.75 & 0.875 & \dots \\
0.75 & 0.875 & 0.64 & \dots \\
\vdots & \vdots & \vdots & $\ddots $
 \end{tabular} \right) \en
   Since these interaction strengths differ significantly more than the ones of different internal states, much higher sensitivity to the distribution of atoms in different vibrational states is achieved, which allows a  faster population-dependent transfer. More fundamentally, however, the difference enables the algorithmic manipulation discussed in the next section.

The configuration with $n$ atoms in the  vibrational ground state $\nu=0$ and hyperfine state ($\alpha$), \state{0}{\alpha}, has the interaction energy
\eq E_{n,0} = \frac{n(n-1)}{2} U_{00} , \en
while  the state with $n-1$ atoms in \state{0}{\alpha}, and $1$ atom in hyperfine state ($\beta$) and the second vibrationally excited state $\nu=2$, \state{2}{\beta}, has a different interaction energy
\eq E_{n-1,1} = \frac{(n-1)(n-2)}{2} U_{00} + (n-1) U_{02}.\en
Raman laser beams with the corresponding two-photon detuning and a suitable wave number difference can transfer exactly one atom to
\state{2}{\beta} in all sites with precisely $n$ atoms, while leaving the population unchanged in all other sites. This transfer step is illustrated in Fig.~\ref{fig:Scheme}, for $n=3$. Sweeping a sequence of pulses resonant with the transfer from traps with $n_{max}$, $n_{max}-1, \dots, 2$ atoms, each followed by the removal of the atoms in the hyperfine state ($\beta$), will  transform sites with $1\leq n\leq n_{max}$ into sites with exactly one atom in \state{0}{\alpha}. Choosing $n_{max}$ large enough, we are left with defects only in the initially vacant sites.

\section{Filling vacant sites by merging}
\label{sec:merging}
\subsection{Constraints of unitarity}
To fix the remaining vacancies, we propose to merge each well of the filtered system with auxiliary wells and to perform operations on the combined wells that result in final states with the unit occupancy component significantly increased. We envision this being performed in parallel across the entire system  by merging planes or strings of sites.

Intuitively, an ``OR''-operation on adjacent wells  maximizes the probability to find at least one atom in the ground state of the desired target well.
Given an initial probability for vacancies $\epsilon$, merging $k$ wells into one could yield a final vacancy probability $\epsilon^k$, {\it i.e.}~a dramatic improvement can be achieved. Such an ``OR''-operation, however, is irreversible, and cannot be achieved by unitary dynamics. For example, in order to merge two wells to a target well (here: the left well), the desired mapping should read:
\eq
\ket{1,0}& \rightarrow &\ket{1,0} \\
\ket{0,1}& \rightarrow &\ket{1,0} \nonumber ,
\en
which is clearly non-unitary. At first sight, it is thus not evident how an improvement of the vacancy probability can be achieved without recourse to read-out of the well populations, as proposed in Ref.~\cite{Weiss04:zeroEntropy}.

Here, we present a unitary protocol that is inspired by the principal idea of the ``OR''-operation, while it circumvents the apparent constraint of unitarity by exploiting the larger state-space of three wells and by using particles in one of the wells as ancilla.

We assume that the vacancy probability in each well is $\epsilon$, and show the eight different possible occupations of three wells in Tab.~\ref{tab:confgs}. The middle well is taken to be the target well, which is occupied (or not) before the protocol according to  $n^{\text{i}}_{\text{m}}$. The  occupation of the target well changes after the hypothetical OR-merger ($n^{\text{OR}}_{\text{m}}$) and after our protocol ($n^{\text{f}}_{\text{m}}$).  As we will show in the following, the final occupation  $n^{\text{f}}_{\text{m}}$ can indeed be achieved via unitary operations.  Our protocol improves only the occupation of the middle well for the initial state $\ket{1,0,1}$, {\it i.e.}~it requires both neighboring wells to be occupied for success. Given the population of this initial state of the order of $\epsilon$, the procedure  leads to a significant decrease of the probability of vacancies.

\begin{table}
\begin{tabular}{llrrr}
$\ket{\Psi_{\text{initial}}}$  & $P$          & $n^{\text{i}}_{\text{m}}$ &  $n^{\text{OR}}_{\text{m}}$ &  $n^{\text{f}}_{\text{m}}$ \\ \hline \hline
 $\ket{1,1,1}$           & $(1-\epsilon)^3$ 		&1 &  1 & 1\\ \hline
$\ket{1,1,0}$  &$\epsilon (1-\epsilon)^2$ & 1    & 1 &  1 \\
$\ket{0,1,1}$    &					 & 1  & 1 &  1 \\
$\ket{1,0,1}$    &					 & 0  & 1 &  1 \\\hline
 $\ket{0,0,1}$    & $\epsilon^2 (1-\epsilon)$ & 0  & 1 &  0 \\
$\ket{0,1,0}$    &					 & 1  & 1 &  1 \\
$\ket{1,0,0}$    &					 & 0  & 1 &  0 \\\hline
$\ket{0,0,0}$    &$\epsilon^3$ 			&  0 & 0 & 0
\end{tabular}
\caption{Overview over probabilities for vacancies in the three-well system. The initial occupations of the three wells correspond to the Fock-state $\ket{n_l,n_m,n_r}$, the respective probabilities are then given by $P$. When the target well is (not) occupied initially,  $n^{\text{i}}_{\text{m}}=1~(0)$. After the application of a hypothetical non-unitary ``OR''-operation, the target well occupation is  $n^{\text{OR}}_{\text{m}}$, while our protocol gives  $n^{\text{f}}_{\text{m}}$. }
\label{tab:confgs}
\end{table}

\subsection{Merging protocol}

\subsubsection{Merging of the central and right well}
\begin{figure}[h]
\includegraphics[width=\linewidth]{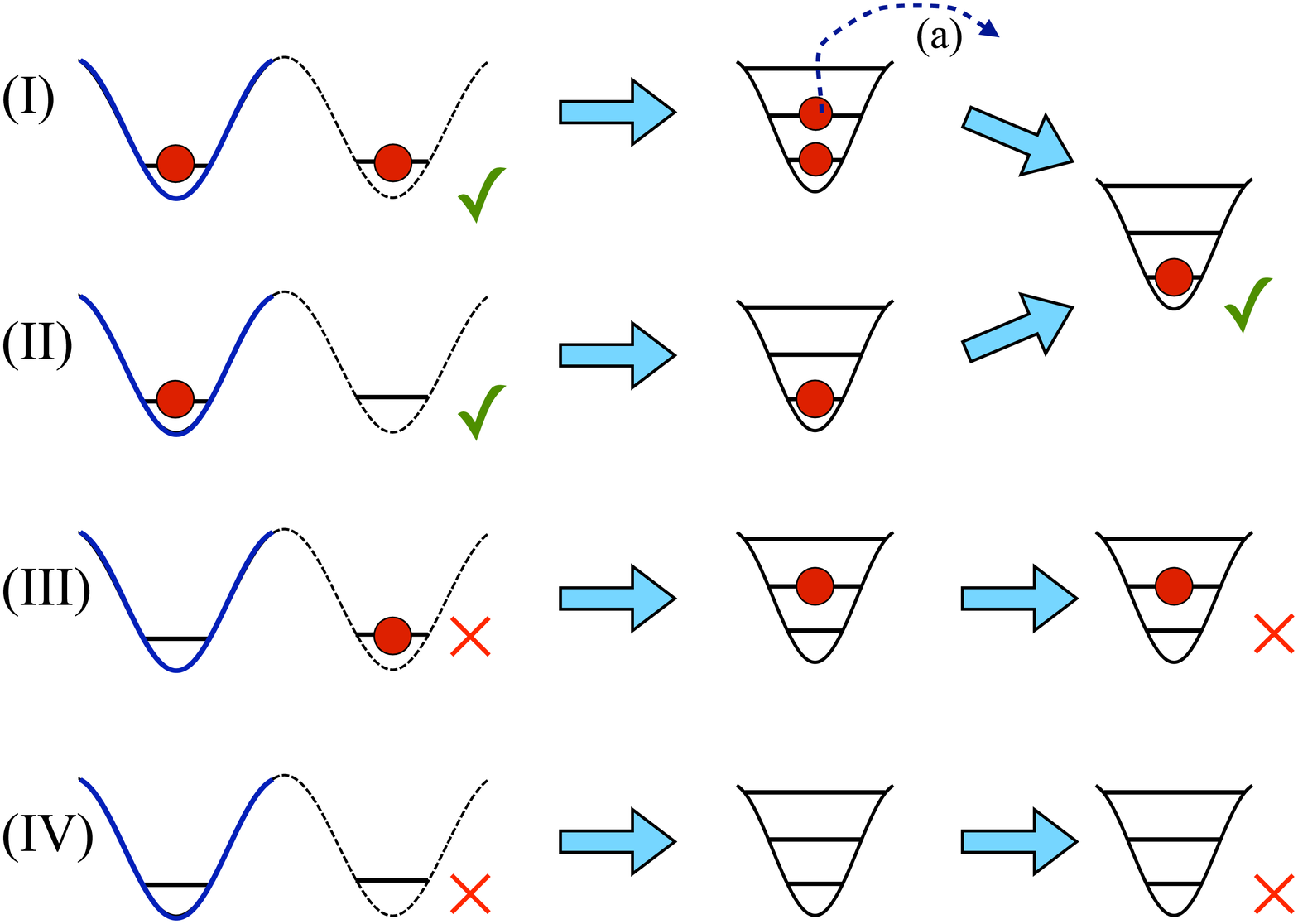}
\caption{(color online) Merging of the left target well (solid blue) with its right neighboring well (dashed), and elimination of doubly occupied wells by removal of the atom in the excited state conditioned on the presence of an atom in the ground state (a). The probability for a singly-occupied ground state is $1-\epsilon$, both before and after the application of the protocol. Wanted (unwanted) configurations are marked with a green crotched (red cross). This part of the protocol in itself does not yet increase the fidelity.}
\label{fig:FirstStep}
\end{figure}

In a first step, we merge the central target well with its right neighboring well, such that the middle (right) well ground state
is mapped to the lowest (first excited) level of the combined well. The process is depicted in Fig.~\ref{fig:FirstStep}, where the third (left) well is omitted for clarity, since this first part of the protocol leaves it untouched.

The merging procedure has to preserve the single-particle states while at the same time being insensitive to the possible interaction between the atoms.
Minimizing the error in the merging process is complicated but in \cite{shersonTransportUnpublished} we  have numerically optimized the merging efficiency and for appropriate parameters show that error probabilities  $\epsilon_m\sim10^{-4}$ can be obtained in less than $100\mu$s even in the presence of interactions. The configuration with two atoms (upper line) can be dealt with by transferring the excited vibrational state population into another hyperfine and vibrational state using  an upper sideband ($\delta\nu=+1$) Raman transition. In principle, this operation would also affect the singly occupied instances (lines (II) and (III) in Fig.~\ref{fig:FirstStep}); however, since the doubly occupied state is initially characterized by the interaction energy $U_{01}$ and in the final state by $U_{02}$ the transition frequency will be shifted by $\Delta U=U_{02}-U_{01}=-0.25~U_{00}$. Since $U_{00}$ is typically 1-20~kHz, this shift is sufficient to ensure a clean spectroscopic discrimination  between the singly and the doubly occupied states. Following the transfer of one of the two atoms to another hyperfine state it is subsequently removed using, {\it e.g.},~a resonant light pulse. This leaves three configurations with a single atom in the well. However, different vibrational levels are now occupied and they cannot be mapped to the ground state by a unitary process. Defining a defect as a missing population of the ground trap level, the error probability is thus still $\epsilon$ and, so far, no improvement in the ground-state occupation has been achieved. The atom in the first excited state ((III) in Fig.~\ref{fig:FirstStep}), can, however, be used as an ancilla particle, as described in  the following.

\subsubsection{Merging of the central and the left well}
\begin{figure}[h]
\includegraphics[width=\linewidth]{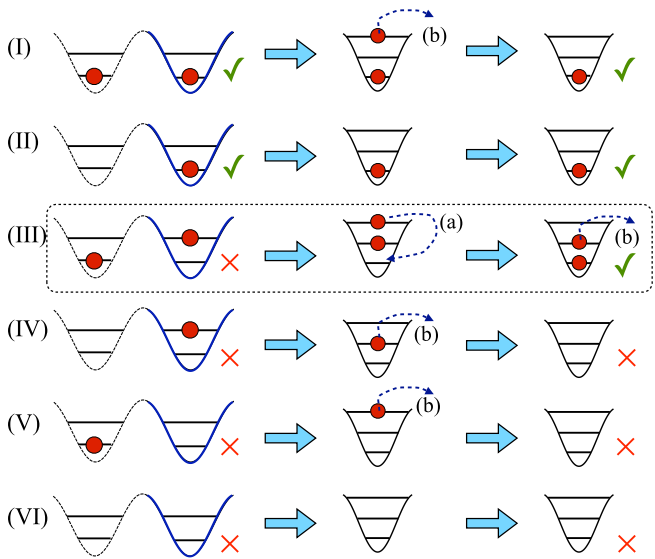}
\caption{(color online) Second step of the protocol: Merging of the target well (solid blue) with its left neighboring well (dashed), which may or may not be singly occupied. The combination of the three distinct outcomes of the first step of the protocol (see the right column in Fig.~\ref{fig:FirstStep}) with the possible occupation or non-occupation of the left well gives rise to six initial configurations, two of them with an occupied target-well ground state (green crotched) and four without occupation of the target state ground state (red crosses). The ground state of the left well is adiabatically merged to become the second excited state of the target well. In (III), the atom in the second excited state is transferred to the ground state, conditioned on the presence of an atom in the first excited state (a). After the transfer, possible population of excited states is removed (b). The final transfer steps on the doubly occupied states result in a singly occupied ground state in {$\text{(I)-(III)}$}. The crucial part of the protocol is thus the conversion of an initial state without occupation of the middle well vibrational ground state to a successful final state with unit occupation.}
\label{fig:thirdwell}
\end{figure}

In a second step, we merge the central target well with its left neighbor, such that the ground state of the left well becomes the second-excited state of the central well. The protocol is shown in Fig.~\ref{fig:thirdwell}, where all possible configurations and their evolutions are depicted.

After the merging, two configurations show double occupancy of the well, either in states \state{0}{\alpha} and \state{2}{\alpha} (I) or in states \state{1}{\alpha} and \state{2}{\alpha} ((III), dashed frame). For the latter, we use the particle in \state{1}{\alpha} as ancilla to transfer the atom in \state{2}{\alpha} to \state{0}{\beta} in a $\delta\nu=-2$ Raman transition, exploiting the energy shifts caused by the state \state{1}{\alpha} population (letter (a) in the figure). To see why this can leave the remaining configurations unchanged, we again need to calculate the difference in interaction energy for the various configurations. For configuration (III) we obtain $\Delta U=U_{00}(1-0.875)=0.125~U_{00}$, whereas the interaction energy of configuration (I) is either unchanged or reduced ($\Delta U \leq 0$) if atoms are transferred to higher bands.

Both two-atom configurations now populate the ground vibrational state with a single atom. Hereafter, all population in excited vibrational levels is mapped to the other hyperfine level $(\beta)$ via a lower sideband $\delta \nu=-1$ Raman transition and subsequently removed (letter (b) in the figure). This lower sideband transition leaves the vibrational ground state population unaffected.

\subsubsection{Reduction of vacancies}
The protocol thus transforms the initial $\ket{1,0,1}$ population (that occurs with probability $\epsilon (1-\epsilon)^2$, see Tab.~\ref{tab:confgs}) into a successful final configuration, such that three of the six configurations in the right column of Fig.~\ref{fig:thirdwell} now have a single atom in the target well ground state. The probability to find a vacancy in the target well thus decreases from its initial value $\epsilon$ to
\eq \epsilon' =\epsilon-\epsilon(1-\epsilon)^2=2\epsilon^2-\epsilon^3 . \en
Clearly, $\epsilon' < \epsilon$, and if $\epsilon$ is small the improvement is dramatic.

The steps can be repeated iteratively, either in a parallel way, such that all groups of three wells are merged simultaneously, or in a protocol in which only the target well is merged with a sequence of pairs of unpurified adjacent wells. We find the recursive update relations after $i>1$ steps,
\begin{equation}
\begin{array}{rl}
  \epsilon_{i,par} &= 2 \epsilon_{i-1,par}^2-\epsilon_{i-1,par}^3\\
  \epsilon_{i,ser} & =2 \epsilon_{i-1,ser} \epsilon-\epsilon_{i-1,ser}\epsilon^2~,
 \end{array}
 \label{eq:epsilonRecusion}
 \end{equation}
for the cases of global parallel and sequential improvements, respectively.

\subsubsection{Unwanted transitions}
During the density-dependent transfer (step (a) in Fig.~\ref{fig:thirdwell}), unwanted transitions can occur, and the atom in the ground state, in lines (I) and (II), may be excited. The probability for this unwanted processes depends on the pulse duration and on the detuning. For a driven two-level transition initially in the ground state, the excited-state population obeys
\eq
P_e(t)=\frac{1}{2}\left(\frac{\chi}{\Omega}\right)^2\left(1-\cos(\Omega t)\right)~,
\label{eq:RabiError}
\en
where $\chi$ is the Rabi frequency, $\Omega=\sqrt{\chi^2+\Delta^2}$ is the generalized Rabi frequency, and $\Delta$ is the detuning from resonance. On resonance this becomes $P_e=\frac{1}{2}\left(1-\cos(\chi t)\right)$ and complete transfer is obtained at $t=\pi/\chi$. For the $\Delta=0.125~U_{00}$ relevant for the merging scheme (a) in Fig.~\ref{fig:thirdwell} the probability of making an unwanted transfer to the other configurations can then be calculated using Eq.~(\ref{eq:RabiError}).

Absorption of lattice photons and collisions with atoms from the background gas fundamentally limit the lifetime $\tau$.  Optimizing under the constraint $\tau=1$~s, we obtain an error of $\epsilon_f\sim 7\cdot10^{-4}(1\cdot10^{-4})$ for $U_{00}/h=1(20)$~kHz at $t=$7(1)~ms pulse duration. Thus, the lattice occupation error may approach, but not go below this value. Still, saturating this bound yields a significant improvement, as we will assess in the next Section.

\subsection{Feasibility}
The merging algorithm can be realized experimentally both with and without single-site resolution. The former is experimentally challenging but can be achieved in a number of different ways with a combination of existing technologies. First, recent advances in high
resolution optical microscopy~\cite{Greiner10:science,Sherson10:MInature}
have opened the possibility to perform the merger either using a focussed tweezer~\cite{rolstonPRA06:singleSite} or with
holographic~\cite{Greiner10:science} or acousto-optic~\cite{Boshier09:AOMpotentials} manipulation of the lattice potential.
Without optical single-site resolution the merger can be realized by the combination of the flipping of the spin of an individual site (or plane) using a magnetic field gradient~\cite{Widera09:MagnAddress,Sherson10:MInature}, spin dependent transport~\cite{Mandel03}, and finally double well merger in a period two superlattice. This approach allows merging to be performed on entire 2D planes in parallel.
Following the first iteration of mergers, the neighboring sites in the auxiliary direction are vacant. For subsequent iterations, atoms have to be brought in over distances that increase linearly with the number of iteration steps. The analysis of  Ref.~\cite{shersonTransportUnpublished} shows that the transport fidelity over several lattice sites does not necessarily decrease with distance. 
Furthermore, since the reduction in state error per iteration is large, as illustrated in Eq.~\ref{eq:epsilonRecusion}, only a few iterations will be needed. 

Alternatively, instead of repeated application of the filtering and merging steps discussed above, one may (adiabatically) merge the ground states of $n$ wells to the $n$ lowest  vibrational states of a single well in a single operation using an appropriate $n$-periodicity superlattice. The simplest realization in a period 3 superlattice consists in manipulations almost identical to the ones sketched out in Fig.~\ref{fig:thirdwell}. Such a lattice has been implemented, {\it e.g.},~in Ref.~\cite{Peil2003}.

\section{Applications and assessment}
\label{sec:applass}
\subsection{Homogeneous Mott insulator} \label{sec:HOMOMI}
To illustrate the accomplishments of the algorithm, we first apply it to a homogeneous MI state. In the zero-tunneling limit we calculate the initial population distribution at a given temperature using a grand canonical ensemble Ansatz for the number distribution:

\begin{equation}
P(n)=\frac{\exp \left(\beta \left[ \uo n - \frac{\Ui}{2}n(n-1) \right] \right)}{Z},
\label{eq:Pn}
\end{equation}

where $\uo$ is the chemical potential, $\beta=1/k_B T$ is the inverse temperature, and
\eq Z=\sum_n \exp{\left( \beta \left[ \uo n - \frac{\Ui}{2}n(n-1) \right] \right)}.\en We see that the distribution depends solely on the two dimensionless parameters
\eq \muU&=&\frac{\uo}{\Ui} , \\
  \TU&=&\frac{k_B T}{\Ui} .  \label{dimensionlesstemp} \en

\begin{figure}[t]
\includegraphics[width=\linewidth]{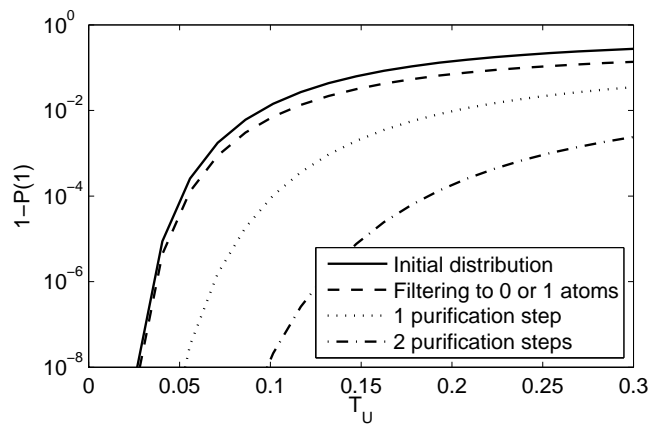}
\caption{Defect probability (probability for vacant or multiply occupied sites) as a function of the dimensionless temperature $T_U$ defined in Eq.~(\ref{dimensionlesstemp})~($\muU=0.5$) in the zero-tunneling MI regime. Initial population calculated according to Eq.~(\ref{eq:Pn}). }
\label{fig:algoPop}
\end{figure}

In Fig.~\ref{fig:algoPop} the solid curve shows the temperature dependence of the probability for defects (occupations other than unity) for a uniform MI with $\muU=0.5$. With sufficient cooling this probability can be brought arbitrarily close to zero, but at currently achievable temperatures ($\TU=0.1-0.2$ \cite{Greiner10:science,Sherson10:MInature}) the error probability remains appreciable. At this chemical potential, applying the filtering operation of Section \ref{numberfiltering} or any other mechanism \cite{0953-4075-43-13-131001,Rabl03,Popp06:gndStateCooling,Bakr:2011fk} only slightly reduces the error probability: Vacancies rather than defects constitute the main source of error. In contrast, a single iteration of our (ideal) filtering and merging algorithm (as described in Section \ref{sec:merging}) offers more than an order of magnitude improvement as shown by the dotted curve. For example, for an achievable temperature of $\TU=0.1$, the initial error probability of $10^{-2}$ is reduced to $10^{-4}$, where the largest contribution of the improvement stems from the merging algorithm. Each subsequent iteration reduces the error further according to Eq.~(\ref{eq:epsilonRecusion}).

\subsection{Inhomogeneous Mott insulator}
In most optical lattice experiments, the atoms experience also a weak harmonic confinement  \eq \Vh(r)=\frac{1}{2}m \otrap^2r^2 .\en  In the combined potential, the atoms have an increasing density towards the center of the  harmonic trap, with MI domains  separated by superfluid regions \cite{PhysRevA.71.063601}. To investigate how our algorithm affects the global density distribution in this case we choose experimentally motivated parameters for $^{87}\mathrm{Rb}$ $\otrap=2\pi\cdot 80$Hz, $\Ui=1.0$kHz,  lattice spacing $d_{Lat}=0.5\mu$m~\cite{Sherson10:MInature}. At each radius $r$ we define a local chemical potential $\uloc(r)=\uo-\Vh(r)$ and apply it to Eq.~(\ref{eq:Pn}) to calculate the position-dependent number distribution  $P(n,r)$.

\begin{figure}[t]
\includegraphics[width=\linewidth]{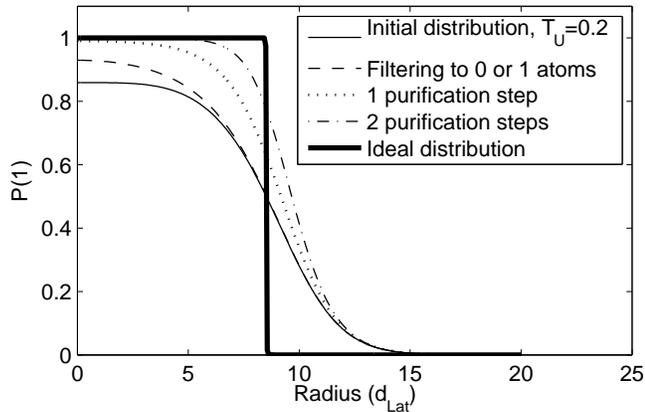}
\caption{Probability for unit occupation as a function of the radius, for  $\muU=0.5$. In the initial distribution (solid thin line), both, vacancies  and multiply occupied sites are not negligible. The filtering operation brings higher occupied sites to unit occupation, which increases the probability for the latter (dashed line). Remaining vacancies are partially eliminated by the application of one (dotted line) or two (dashed-dotted line) merging steps, which increases the probability for unit occupation and leads to a large close-to-ideal MI domain. }
\label{fig:algoVsR}
\end{figure}

\begin{figure}[t]
\includegraphics[width=\linewidth]{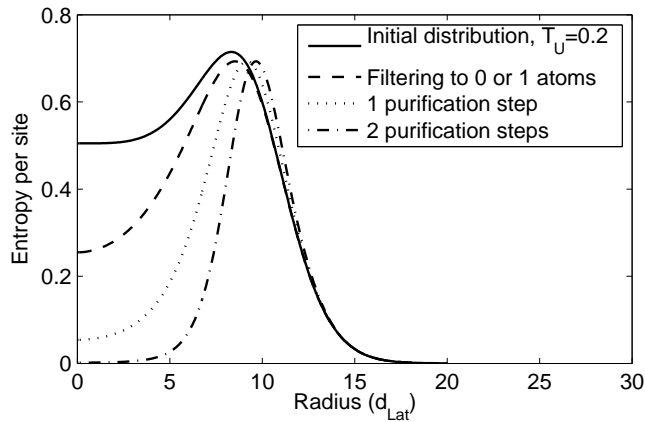}
\caption{Entropy per site, as a function of the radius, calculated according to Eq.~(\ref{EntropyPerSite}). While the filtering and the merging steps dramatically reduce the on-site entropy for low radii, the low average particle number per site prevents this for larger radii.}
\label{fig:entro}
\end{figure}

 For $\TU=0.2$ and $\muU=0.5$, we obtain the probability for unit occupancy plotted as the thin solid curve in Fig.~\ref{fig:algoVsR}.
 Clearly, due to the finite temperature, a significant part of the sites contain other than a single atom. After filtering to 0 or 1 atoms per site (dashed line), only vacancies remain as an error source. The application of a merging step then removes the vacancies and ensures a uniformly filled lattice, which is extended further by the second merging step. A very pure domain with $n=1$  is already formed after the first iteration and the size of this domain grows with subsequent iterations. Rather than the sharp transition of the $T=0$ MI (thick solid line), the pure domain is, however, surrounded by a region in which sites have, both, an appreciable probability to be occupied and to be unoccupied. In other words, this region contains  a high entropy,
 \eq
 H(r)= - \sum_{n=0,1} P(n,r) \log P(n,r) , \label{EntropyPerSite}
 \en
which is also depicted in Fig.~\ref{fig:entro}.

For applications in quantum computation this high-entropy shell may be of no importance as long as the inner region is sufficiently pure. However, for application in quantum simulation of spin Hamiltonians it will be crucial to realize a state with a high overlap with the full quantum many-body ground state. This can be achieved by using a skimmer to remove all atoms beyond a certain cut-off radius. Such an operation can either be realized with an optical tweezer focused on a single lattice site~\cite{weitenberg:address11} or with magnetic single-site spectroscopic resolution~\cite{Widera09:MagnAddress,Sherson10:MInature} in a cylindrically symmetric magnetic field configuration. We quantify the purity of the states of the system as the overlap (OL) with the ideal many body state with precisely one atom in every lattice well, $i$, of a system with radius $\rcut$,
\eq \text{OL}=\prod_{r_i <\rcut} P_i(1). \en
A finite value of this overlap ensures that one has a finite component of the relevant state in, {\it e.g.}, a quantum simulation or computation.

\begin{figure}[ht]
\includegraphics[width=\linewidth]{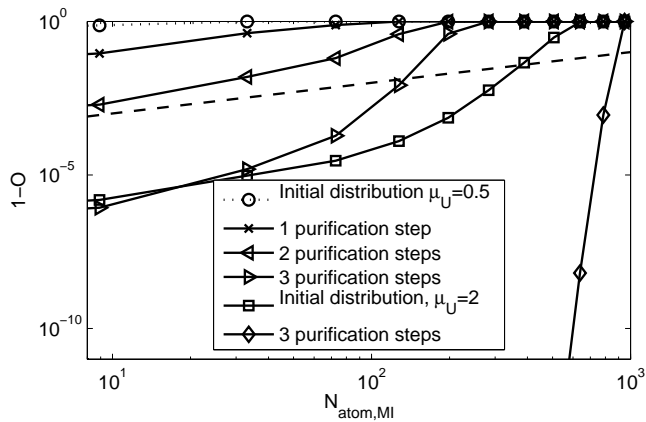}
\caption{Infidelity of the states with all population beyond a critical radius removed  shown as a function of the system size. Initially $\TU=0.2$. The dashed line indicates the minimum infidelity imposed by $1\cdot10^{-4}$ errors in the filtering and merging operations.}
\label{fig:FidelvsCutRadius}
\end{figure}

In Fig.~\ref{fig:FidelvsCutRadius} we plot the infidelity, $1-\text{OL}$, versus the atom number in a perfect 2D MI state of size $\rcut$.
The upper four curves at $N=100$ illustrate how in the low filling case ($\muU=0.5$) a high-fidelity region is formed upon repeated application of our algorithm. For three iterations a state with around 100 atoms and an overlap better than $0.99$ is realized taking into account filtering and merging errors (thick line), {\it i.e.}~a minimal defect probability of $\epsilon_f=1\cdot 10^{-4}$ is assumed. In the last two curves we illustrate the intrinsic power of the filtering operation at high filling and low enough temperature. With squares we show the result of the filtering on a $\muU=2$ state with initial temperature $\TU=0.2$. As can be seen, a high-fidelity state of around 200 atoms is formed. Three iterations of our algorithm yield a nearly $10^3$ atom system with very high fidelity: the core of an impressive large scale quantum computer.

In the case of fermions, Eq.~(\ref{eq:Pn}) contains only zero or one atom in the ground state but our initial filtering has to be extended to incorporate the possibility of zero and unit population in the vibrationally excited states. Such filtering is simply implemented by $\delta \nu=-1$ Raman transitions and removal of the final internal $(\beta)$ state atoms, which leaves the vibrational ground state $(\alpha)$ population intact. The merging protocol of Section \ref{sec:merging} relies on interactions, which vanish for fermions occupying the same hyperfine state $(\alpha)$. Interactions between the different hyperfine states of the final state can be present and  entropy removal by merging and occupation dependent filtering can be realized also in the fermionic case.

\FloatBarrier
\section{Conclusions} \label{sec:conclusions}
In conclusion, we have presented an efficient way to realize density-dependent manipulation of optical lattice sites using the dependence of the interaction energy on the vibrational states occupied by the atoms. This can be used to filter all higher occupations to unity and thus remove entropy from the system. However, filtering does not solve the fundamental problem of vacancies, which remains important for low temperatures. We address this problem by adiabatically merging the contents of several lattice sites, followed by filtering, adapted to the vibrational states occupied after the merging. Our procedure although experimentally challenging can be realized with a combination of existing technology. It is based on global operations, and does not necessarily require single-site addressing, as in other proposals \cite{Weiss04:zeroEntropy}. As we have shown in Section \ref{sec:HOMOMI}, a considerable increase in fidelity is achieved by the application of the  merging procedure.

The focus of our work has been on identifying efficient and fast operations, which we believe provide an important step towards the creation of near-zero entropy states of atoms in optical lattices that are useful for quantum simulations and large-scale quantum computation. Finally, we note that the sensitivity to atomic occupation number using vibrational excitations may find applications in the control of many-body quantum states of atoms in lattices and microtraps, {\it e.g.}, by inducing energy shifts of selected Fock components and thus tailoring completely new varieties of Bose-Hubbard-like Hamiltonians. \\
\vspace{1cm}

\subsection*{Acknowledgements} The authors acknowledge conversations with Uffe Poulsen and Nicolai Nygaard. JFS acknowledges funding from EU (FP7-PEOPLE-2007-2-1-IEF, No:219560) and the Danish National Research Council (FNU-STENO).

\end{document}